\documentclass[english]{article}
\usepackage[T1]{fontenc}
\usepackage[latin1]{inputenc}
\usepackage{graphicx}

\makeatletter

\newcommand{\lyxaddress}[1]{
\par {\raggedright #1
\vspace{1.4em}
\noindent\par}
}

\usepackage{babel}
\makeatother
\begin{document}

\title{Anisotropic colloids through non-trivial buckling.}

\author{C. Quilliet$^{1,2}$, C. Zoldesi$^{2}$, C. Riera$^{3}$, A. van
Blaaderen$^{2}$, A. Imhof$^{2}$}

\maketitle

\lyxaddress{$^{1}$ Laboratoire de Spectrométrie Physique, CNRS UMR 5588 \& Université
Joseph Fourier, 140 avenue de la Physique, 38402 Saint-Martin d'Hères
Cedex, France}

\lyxaddress{$^{2}$Soft Condensed Matter, Debye Institute, Faculty of Science,
Utrecht University, Princetonplein 5, 3584 CC Utrecht, The Netherlands}

\lyxaddress{$^{3}$DEAS, Harvard University, 29, Oxford Street Cambridge MA 02138,
USA}

\begin{abstract}
We present a study on buckling of colloidal particles, including experimental,
theoretical and numerical developments. Oil-filled thin shells prepared
by emulsion templating show buckling in mixtures of water and ethanol,
due to dissolution of the core in the external medium. This leads
to conformations with a single depression, either axisymmetric or
polygonal depending on the geometrical features of the shells. These
conformations could be theoretically and/or numerically reproduced
in a model of homogeneous spherical thin shells with bending and stretching
elasticity, submitted to an isotropic external pressure.
\end{abstract}
Pacs : 46.70.De (Static buckling and instability), 82.70.Dd (Colloids),
89.75.Kd (Patterns)

\section{Introduction}

Anisotropic colloidal particles made using spheres have been the subject
of various studies in recent years. These types of colloids can be
obtained in very different ways, for example either by deformation
of spherical particles\cite{FerNi2005,Shkly2005,vanDillen2003}, or
by forming clusters of them\cite{Yin2001,Mano2003,Johnson2005,Zerrouki2006}.
Such objects are good candidates to generate anisotropic colloidal
crystals. Photonic bandgap (PBG) calculations showed that such crystals
should improve the expected performances (larger bandgap, more convenient
wavelengths)\cite{Yin2001,vanDillen2004,Liddel2003}. Colloids with
a hollow interior (spherical shells) are particularly interesting
also for their mechanical properties, which make them potentially
important for a variety of applications, such as drug delivery, catalysis
and biotechnology, and when filled with gas as contrast agents for
ultrasound or echographic imaging\cite{Nanda1993,Marmottant2005}.
Due to the relevant engineering situations as well as the biomechanics,
the problem of the deformation of a spherical shell under external
constraints has been recently investigated, both experimentally and
numerically\cite{Vinogradova2004b,Leporatti 2000,Dubreuil2003,Ivanovska2004,Gao2001,Michel2006,Fery2007}.

In this study, we present anisotropic colloids obtained by buckling
of spherical shells. The buckling was induced by dissolving or evaporating
the solvent enclosed in the slightly porous shells. This causes a
stress comparable to an isotropic pressure on a spherical airproof
shell\cite{Zoldesi2007}. After postbuckling, the colloidal shells
show bowl-like conformations, either with axisymmetric or polygonal
symmetry. We observed them using both transmission optical microscopy
and transmission electron microscopy (TEM), and compared them with
configurations obtained from Surface Evolver simulations, using a
model of homogeneous elastic (bending and in-plane stretching) spherical
shells submitted to an isotropic external pressure.

\section{Buckling under evaporation and in solution\label{sec:Experimental-features-:}}

\subsection{Methods}

Spherical colloidal shells were prepared following previous work by
Zoldesi and co-authors\cite{Zoldesi2005,Zoldesi2006}. They vigorously
mixed dimethyldiethoxysilane (DMDES) and an aqueous solution of ammonia
($NH_{3}$), providing a very monodisperse and stable oil-in-water
emulsion with droplets of micrometric size. The oil consists of low
molecular polydimethylsiloxane (PDMS) oligomers. By adding tetraethoxysilane
(TEOS), a solid shell forms at the surface of the droplets, consisting
mainly of PDMS with average oligomer length 4, crossed-linked with
hydrolyzed TEOS units. As shown in ref \cite{Zoldesi2006}, the shells
are porous and then allow small molecules to pass through. As the
low molecular PDMS can be dissolved in ethanol \cite{Lee2003}, adding
ethanol (an equal volume in the present study) to such an aqueous
suspension of shells filled with oil, leads to dissolution of the
encapsulated oil into the external medium. For thick enough shells
(\char`\"{}capsules\char`\"{} and spheres in the nomenclature of ref
\cite{Zoldesi2006}, i.e. shell thickness over 100 nm), this leads
to a suspension of solvent-filled spherical particles. These particles
are then sedimented by centrifugation and redispersed in ethanol.
They may be dried afterwards and observed by electron microscopy,
exhibiting shape modifications ({}``buckling under evaporation'')
or not \cite{Zoldesi2005,Zoldesi2006}, depending on the ratio of
their shell thickness to radius \cite{Zoldesi2007}.

Thinner shells, however, will already buckle in solution when ethanol
is added. Thin shells consist of much larger PDMS oil droplets coated
with a very thin solid organosilica layer. They were prepared using
the same procedure as described in ref.\cite{Zoldesi2005} for {}``microballoon''
particles. In order to obtain bigger particles, we increased the concentration
of DMDES and ammonia up to 5\% v/v, and the droplets were allowed
to grow for three days before the encapsulation step. This resulted
in somewhat more polydisperse particles with diameters between 3 and
6 $\mu$m. Since this small polydispersity prevented from determining
the shell thickness through static light scattering, we had to use
alternative techniques to estimate it. A range of 5-20 nm was proposed
\cite{Zoldesi2005} by considering that all the TEOS forms a dense
silica shell, but careful observations of transmission electronic
microscopy (TEM) pictures suggest 10-40 nm, which is likely since
PDMS oligomers are known to co-polymerize with TEOS, hence contributing
to the shell thickness.

\subsection{Experimental results\label{sec:Post-buckling-conformations.}.}

\begin{figure}
\includegraphics[width=50mm,keepaspectratio]{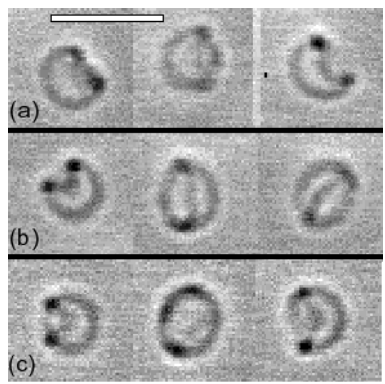}

\caption{\label{fig:Typical-transmission-optical}Typical transmission optical
microscopy images of particles from a suspension of colloidal spheres
filled with oil, in a mixture of water and ethanol. Three different
objects are displayed under three different views; they all show the
same coffee-bean buckled shape, with an elongated depression. Scale
bar 4 $\mu m$.}
\end{figure}

In order to resolve the post-buckling structure by transmission optical
microscopy, we used batches of larger particles with radii R between
2 and 3 $\mu$m. Once buckled in solution through addition of ethanol
as described in the previous section, these particles hold a single
depression with significant volume compared to that of the initial
sphere. Furthermore, in some cases the depression is not axisymmetric
anymore. In one batch (fig. \ref{fig:Typical-transmission-optical}),
we did observe apparently identical objects, all with an elongated
depression that gives the whole object a coffee-bean shape. This type
of conformations was reported in literature for red blood cells \cite{Bessis1956}
and for polystyrene shells previously filled with organic compounds
and evaporated in air \cite{Okubo2001,Okubo2003}, and was reproduced
by numerical simulations\cite{Lim2002}.%

\begin{figure}
\includegraphics[width=40mm,keepaspectratio]{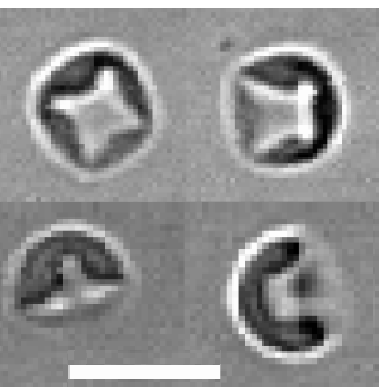}\includegraphics[width=40mm,keepaspectratio]{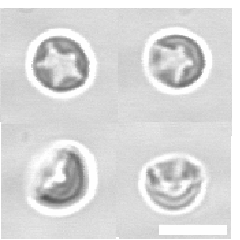}\includegraphics[width=18mm,keepaspectratio]{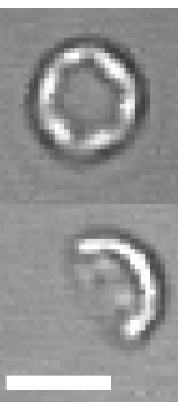}\includegraphics[width=18mm,keepaspectratio]{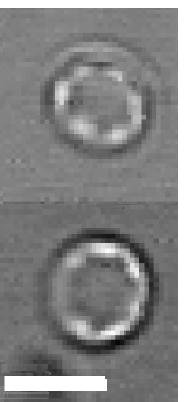}\includegraphics[width=18mm,keepaspectratio]{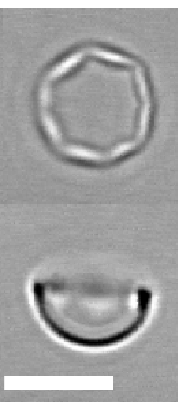}

\caption{\label{fig:Spherical-shells-buckled}Spherical shells buckled in
solution, showing 4 to 8 wrinkles (a-e). Each sub-figure shows different
transmission optical microscopy views of the same object. Scale bar
5$\mu m$, except subfigure c: 2$\mu m$.}
\end{figure}

In other cases, the depression presented a polygonal aspect due to
regularly spaced radial wrinkles, in a number varying from 4 to at
least 8 (fig. \ref{fig:Spherical-shells-buckled}), 7-8 being the
upper limit that we could still distinguish and count for these shell
sizes with optical microscopy. 

The obtainment of such structures in solution from synthetic colloids
is a total novelty. Besides, shapes with a depression presenting a
3-fold symmetry were previously observed in red blood cells\cite{Bessis1956,Lim2002,Okubo2001}
or in dried polymer particles \cite{Okubo2001,Okubo2003} but we could
not find in the literature observations concerning a higher number
of wrinkles, on any system. To our knowledge, theoretical predictions
leading to such wrinkles do not exist either, except when generated
by a point \cite{Fitch68,Pauchard1998} or a flat \cite{Kitching75}
load, which is not the case here.

It is interesting to find out whether some shell heterogeneity has
to be invoked to explain such a non-symmetry, or if a model of elastic
homogeneous spherical shells is sufficient to recover these nontrivial
shapes. For this purpose, we are interested in a model of thin spherical
shells with bending and 2D stretch elasticity, that we derived both
analytically and numerically. Such an approach was likely to give
hints on features hardly accessible by experiments, such as, here,
shell thickness, or the successive steps that lead to the final (and
observable) shape adopted by the shells.

\section{Theoretical part: thin plate elasticity\label{sub:Simulation-of-elastic}\label{sub:Thin-plate-elasticity}}

The elastic energy stored in the deformation of a thin sheet of an
isotropic and homogeneous material may be split into a bending and
a stretching part \cite{BenAmar1997}, and both can be written in
terms of surface elasticity:\begin{equation}
F=\int_{shell\ surface}(\frac{1}{2}\kappa\left(c-c_{0}\right)^{2}+\frac{1}{2}\epsilon_{ij}K_{ijkl}\epsilon_{kl})dS\label{eq:Elastic energy}\end{equation}
where $\kappa$ is the bending constant, $c_{0}$ the spontaneous
curvature of the shell (which is zero for an unstress flat sheet,
but 1/R for a spherical shell without constraints) \cite{Helfrich1973},
and $\epsilon_{ij}$ and $K_{ijkl}$ respectively the two-dimensional
strain and elasticity tensors. The nonzero terms of the two-dimensional
elasticity tensor are $K_{xxxx}=K_{yyyy}=\frac{A}{1-\nu^{2}}$, $K_{xxyy}=K_{yyxx}=\frac{\nu A}{1-\nu^{2}}$
and $K_{xyxy}=K_{yxyx}=\frac{A}{1+\nu}$ (with $A$ the equivalent
two-dimensional Young modulus and $\nu$ the equivalent two-dimensional
Poisson ratio)\cite{Landau}. The stretch elasticity term can thus
be rewritten as $\frac{A}{2(1+\nu)}\left[\mbox{Tr}(\epsilon^{2})+\frac{\nu(\mbox{Tr}\epsilon)^{2}}{1-\nu}\right]$.
We ignored the gaussian curvature term\cite{Pauchard1997} since,
according to the Gauss-Bonnet theorem, its integral depends only on
the topology for a closed surface.

To establish the link between these two-dimensional elastic parameters
and the three-dimensional features (sheet thickness $d$, Young modulus$E$
and Poisson ratio $\sigma$ of the bulk material), we follow Landau's
approach\cite{Landau}. For what concerns bending, the distribution
of stress in a thin plate under flexion at equilibrium leads:

\begin{equation}
\kappa=\frac{Ed^{3}}{12(1-\sigma^{2})}\label{eq:StretchModulusX2}\end{equation}

The 2D Young's modulus $A$ was not taken equal to $Ed$ because this
would correspond to a {}``planar'' deformation in Landau's terminology,
\emph{i.e.} a deformation of plates at constant thickness, which seems
not adequate here. The homogeneous {}``longitudinal'' deformation,
\emph{i.e.} without constraints in the perpendicular direction, is
indeed more adapted to what happens in the shells. This leads:

\[
A=\frac{1+2\sigma}{(1+\sigma)^{2}}Ed\]
 \[
\nu=\frac{\sigma}{1+\sigma}\]

It is worthwhile to notice that, as the three-dimensional Poisson
ratio $\sigma$ has a maximum value of 0.5 (incompressible materials),
the two-dimensional Poisson ratio $\nu$ of such free plates cannot
exceed 1/3. This point was neglected in previous work \cite{Tsapis2005},
and can become of some importance if one wants to make a link between
the parameters choosen for simulation and the geometrical properties
of the shells.

As the surface integral scales like $R^{2}$, the dimensionless Föppl-von
Karman number $\gamma=\frac{AR^{2}}{\kappa}$ is likely to drive the
succession of configurations resulting from the balance between bending
and stretching. In this model of thin shell of elastic isotropic material,
we then expect:

\begin{equation}
\gamma=12(1-\frac{2\sigma^{2}}{1+\sigma})\left(\frac{R}{d}\right)^{2}=12(1-\frac{2\nu^{2}}{1-\nu})\left(\frac{R}{d}\right)^{2}\label{eq:DefFvK_lambda}\end{equation}

It is interesting to note that this model predicts conformations to
be independent of $E$, and to finally depend only on the relative
thickness $d/R$ and the Poisson ratio.

In such a model, we can calculate the elastic energy of an initially
unstrained spherical surface which inner volume decreases by $\Delta V$
its initial value $V$, in two conformations: when the sphere remains
spherical, and following reference \cite{Quilliet2006} when an axisymmetric
depression is created by inverting a spherical cap:

\begin{equation}
U_{sphere}=4\pi R^{2}\times\frac{A}{9(1-\nu)}\left(\frac{\Delta V}{V}\right)^{2}\label{eq:Usphere}\end{equation}

\begin{equation}
U_{axisym}=\pi\,\frac{AR^{2}}{\gamma}\left(\frac{d}{R}\right)^{-\frac{1}{2}}\left[\sin\alpha\left(\tan\alpha-\left(\frac{d}{R}\right)^{1/2}\right)^{2}+4\left(\frac{d}{R}\right)^{1/2}\left(1-\cos\alpha\right)\right]\label{eq:Ucapsule}\end{equation}

where $\frac{d}{R}$ can be expressed as a function of $\gamma$ and
$\nu$ through equation \ref{eq:DefFvK_lambda}. Parameter $\alpha$
is the half-angle of the revolution cone apexed at the sphere center,
and in which the axisymmetric depression inscribes. This half-angle
relates to the relative volume variation through:\begin{equation}
\frac{\Delta V}{V}=\frac{1}{2}(1-\cos\alpha)^{2}(2+\cos\alpha)\label{eq:deltaV_vs_alpha}\end{equation}

In the limit of very thin shells and small volume variations, one
can show that

\[
U_{axisym}\approx\frac{\pi}{12^{1/4}}AR^{2}\gamma^{-\frac{3}{4}}\left[1-\frac{2\nu^{2}}{1-\nu}\right]^{-\frac{1}{4}}\alpha^{3}\]

Or, as a function of the relative volume variation:

\[
U_{axisym}\approx\frac{4}{3}\frac{\pi}{2^{1/4}}AR^{2}\gamma^{-\frac{3}{4}}\left[1-\frac{2\nu^{2}}{1-\nu}\right]^{-\frac{1}{4}}\left(\frac{\Delta V}{V}\right)^{3/4}\]

In this limit, $U_{sphere}=U_{axisym}$ would then happen for relative
volume variations:\begin{equation}
\frac{\Delta V}{V}\propto\gamma^{-\frac{3}{5}}\label{eq:PowerLawBuckling2}\end{equation}

which provides a scaling law for the {}``sphere towards capsule''
buckling.

These theoretical calculations will be compared to simulation results
in paragraph \ref{Simus_vs_theo}.

\section{Simulations}

\subsection{Modus operandi}

The simulated configurations presented hereafter were obtained using
the free software Surface Evolver \cite{Brakke1992}, in which the
elastic energy given by equation (\ref{eq:Elastic energy}) is minimized
in the space of conformations. The stretch energy term is in fact
calculated using the Cauchy-Green strain tensor, which is accurate
for describing deformations of larger amplitude. The minimization
was performed by alternating gradient, conjugate gradient and hessian
methods. Stochasticity was introduced by jiggling the position of
the vertices at the beginning of each minimization (\emph{i.e.} at
each volume step when the volume is decreased by steps). We tested
that the number of vertices is high enough to avoid an influence of
the mesh on the conformations. Furthermore, we checked that the symmetry
of the mesh has no influence on the position of the wrinkles, by using
an isotropic randomized mesh. Such an approach was initiated by Tsapis
et al \cite{Tsapis2005}.

We explored a discrete range of $\gamma$, and restricted our simulations
to the case of an incompressible material, \emph{i.e.} $\nu=1/3$
($\sigma=1/2)$.

\begin{figure}
\includegraphics[width=42mm,keepaspectratio]{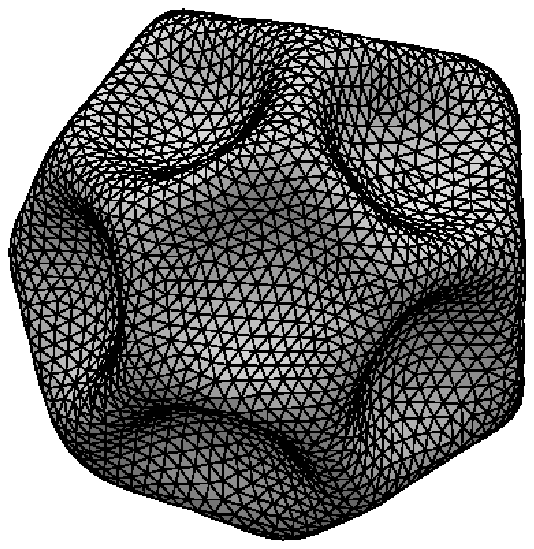}\includegraphics[width=36mm,keepaspectratio]{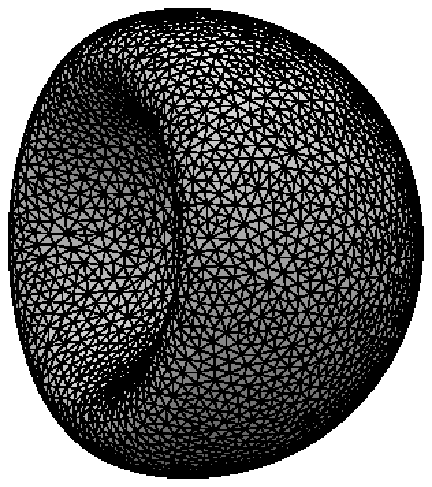}

\caption{\label{fig:Simulation-obtained-for}Conformations numerically obtained
with Surface Evolver for $\gamma$=743, $\nu$=0.333 (which corresponds
to a shell of relative thickness $d/R$=0.116 of an incompressible
material with $\sigma$=0.5),$\frac{\Delta V}{V}$=0.171, $c_{0}=1/R$,
3594 vertices. Left: simulation performed by directly minimizing at
$c_{0}=1/R$ for each decreasing volume step: N=17 depressions. Right
: simulation performed first with $c_{0}=0,$ then $c_{0}=1/R$. The
elastic energy is 3.10 times larger for the {}``potato'' (left)
than for the capsule (right).}
\end{figure}

Given these elastic parameters, a first set of minimization was performed
through stepwise decrease of the inner volume of an initially spherical
surface, and stress free (\emph{i.e.} $c_{0}=1/R$), with a minimization
at each volume step. This leads to roughly isotropic structures with
depressions regularly spaced on the surface, such as in reference
\cite{Tsapis2005} or on figure \ref{fig:Simulation-obtained-for}-a.
The number of depressions is found to increase with $\gamma$.

Different shapes of much lower energy (for the same elastic parameters)
could be obtained through more sophisticated minimizations. This was
done by reversibly acting on the spontaneous curvature $c_{0}$ of
the shells. Since the shells are formed by templating on the oil droplets,
one can assume that they are unstrained in their initial state and
$c_{0}$ is expected to be $c_{0}=1/R$. But when $c_{0}$ is changed
to zero , conformations qualitatively different could be reached:
one obtains {}``capsules'' with a single axisymmetric depression.
Minimizing again with $c_{0}$ back to $1/R$ preserves this capsule
conformation, with an energy lower than the potato shape, as exemplified
on figure \ref{fig:Simulation-obtained-for}-b. Temporarily imposing
a zero spontaneous curvature is likely to lower the energy barrier
for the merging of two different depressions at the surface of the
sphere, since merging happens through flattening of the high positive
curvature ridge that separates the two depressions. This trick apparently
helps to get out of some local minimas in which the simulated conformations
are easily quenched, as is quite usual in buckling problems.

In the following, such a zero curvature cycle was systematically performed
at each volume step, in order to facilitate conformation changes.

\subsection{Results}

We performed simulations of an elastic closed surface, initially spherical
and unstrained, which inner volume is decreased by volume steps $dV=0.0190\times V_{0}$,
$V_{0}$ being the volume of the initial sphere. In all the experiments,
$V_{0}$ and $\kappa$ were kept unchanged and $\gamma$ was varied
from 271 to 29160 by changing $A$. 

Simulations are stopped when the surface interpenetrates, which happens
for inner volumes of the order 7-11\% the initial volume $V_{0}$.

Up to $\gamma$=583, the volume decrease causes a buckling toward
the {}``capsule'' conformation, \emph{i.e.} with a single axisymmetric
depression, until the surface interpenetrates.

From $\gamma$=933, axisymmetric capsules undergo a second transition
when the volume goes on decreasing, toward a non-axisymmetric conformation.
The onset of this second transition is harder to detect since the
depression, except for the highest values of $\gamma$, only slowly
evolves toward a polygonal shape, and then the corners of the polygon
turn into wrinkles (limits of the depression concave between the apices).
Quantitative data concerning buckling thresholds will nevertheless
be given in the paragraph \ref{Simus_vs_theo}. We could this way
obtain such wrinkled bowls with 5 wrinkles or more. Some trials with
re-increasing volume from a wrinkled state also provided conformations
with 3 and 4 wrinkles.

\begin{figure}
\includegraphics[scale=0.65]{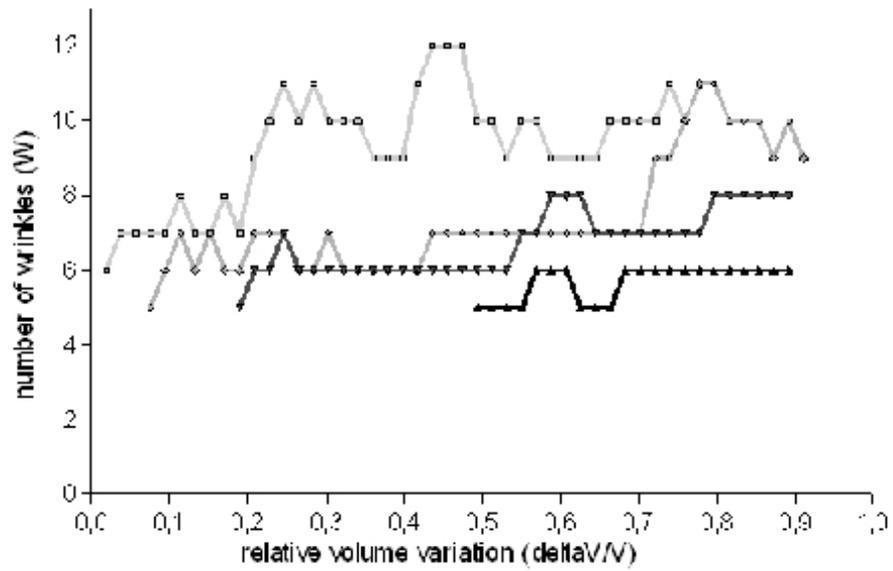}

\caption{\label{fig:W_vs_DVsurV}Typical evolution of the number W of wrinkles
held by the single depression after secondary buckling (simulations
performed for $\gamma$=2333 (light grey, squares), 7465 (medium grey,
diamonds), 15163 (dark grey, downwards pointing triangles) and 29160
(black, upwards pointing triangles; $\nu$=0.333). }
\end{figure}

More quantitatively, figure \ref{fig:W_vs_DVsurV} provides the number
$W$ of wrinkles observed throughout a stepwise volume decrease for
simulations with different $\gamma$ values. One can notice that there
is some fluctuation on $W$, of one, or even two, units. In fact,
performing decreasing volume simulations with the same parameters
does not always lead to the same number of wrinkles. We could get
in these situations an order of magnitude for the energy difference
between two close conformation (i.e. $W=\pm1$): it can be as small
as a few tenths of percent. The energy of a conformation is not strictly
determining its occurence: the path followed in the space of conformations
has some importance. This is why we restricted most of our study to
a sequence of minimization that is likely to reproduce the experimental
situation, \emph{i.e.} decreasing the volume step by step.

Figure \ref{fig:W_vs_DVsurV} shows a weak tendency for the number
of wrinkles $W$ to increase with the relative volume variation $\frac{\Delta V}{V}$
once the buckling has occured, in the same way that was observed in
macroscopic indentation experiments\cite{Pauchard1998}.

\begin{figure}
\includegraphics[scale=0.9]{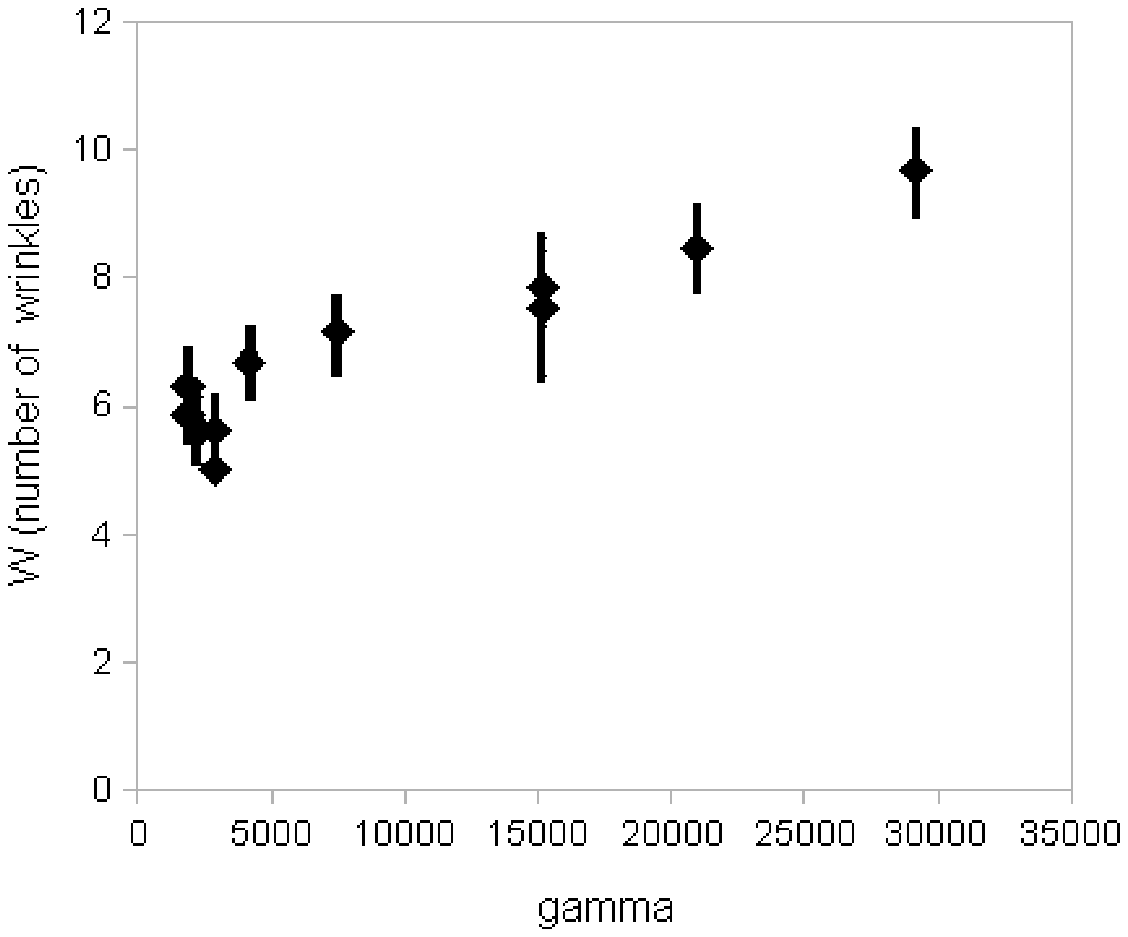}

\caption{\label{fig:WvsGamma}Number $W$ of wrinkles held by the single depression
after secondary buckling, averaged between $\frac{\Delta V}{V}$ =
0.53 and $\frac{\Delta V}{V}$ = 0.76 (error bar is the standard deviation),
as a function of the dimensionless Föppl-von Karman number $\gamma=\frac{AR^{2}}{\kappa}$.}
\end{figure}

More obvious is the variation of $W$ with $\gamma$. In order to
precise a variation of a few units on a discrete quantity, we averaged
$W$ on a range of $\frac{\Delta V}{V}$ where the conformation holds
wrinkles for all the values of $\gamma$, \emph{i.e.} $\frac{\Delta V}{V}$
between 0.53 and 0.76 (figure \ref{fig:WvsGamma}). This puts into
evidence an increase of $W$ with increasing $\gamma$. Wrinkles being
more numerous with decreasing $\frac{d}{R}$ goes in the sense of
intuition: a thinner plate folds more easily, and hence makes more
folding patterns.

\subsection{Comparison between simulations and experiments}

\begin{figure}
\includegraphics[scale=0.5]{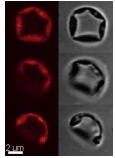}

\caption{\label{fig:CoquillesFluo}Three different views of a buckled spherical
shell labelled with RITC, in confocal fluorescence microscopy (left)
and in transmission microscopy (right). The shell is clearly self-contacting
at its convexe part. }
\end{figure}

The inner volume of the conformations shown in figure \ref{fig:Spherical-shells-buckled}
is not easy to determine precisely. Nevertheless, an important experimental
observation is that buckling deformations never relaxed back toward
the initial spherical shape after complete evaporation or complete
dissolution of the inner oil, yet it means that the last water/air
or oil/water interfaces, that were pulling the shell inwards, have
disappeared. It is then likely that the shrinking brings opposite
surfaces close enough to one another to be sensitive to Van der Waals
attraction, which would stabilize the buckled conformation against
elastic shape recovery after vanishing of the capillary forces. This
hypothesis, of initially opposite parts of the shells that contact
in the conformations experimentally obtained, seems to be confirmed
by confocal pictures of buckled shells (figure \ref{fig:CoquillesFluo}).

In order to compare the shapes obtained through simulations (with
decreasing volume) with the experimental ones, we thus focused on
the shapes obtained just before self-contact.

\begin{figure}
\includegraphics[width=23mm,keepaspectratio]{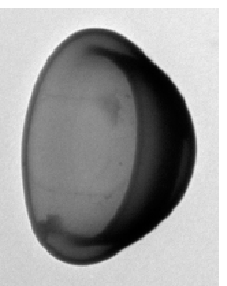}\includegraphics[width=18mm,keepaspectratio]{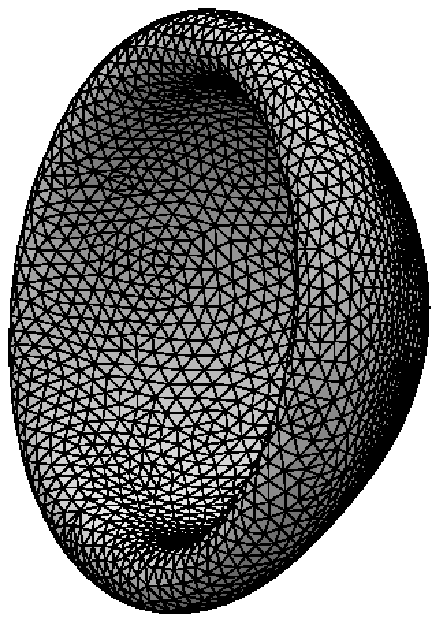}\includegraphics[width=25mm,keepaspectratio]{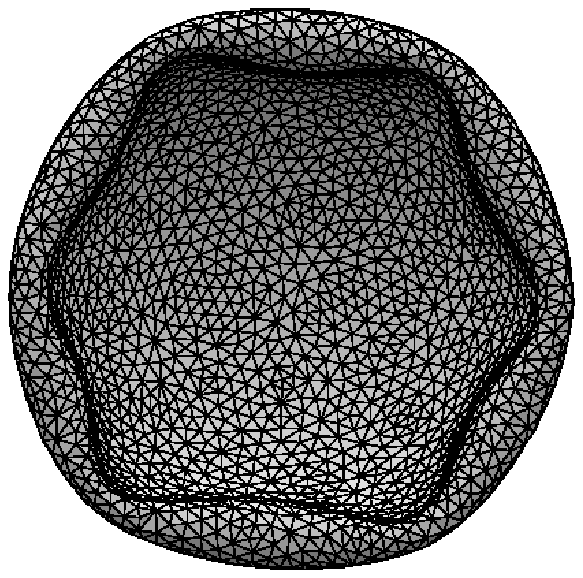}\includegraphics[width=23mm,keepaspectratio]{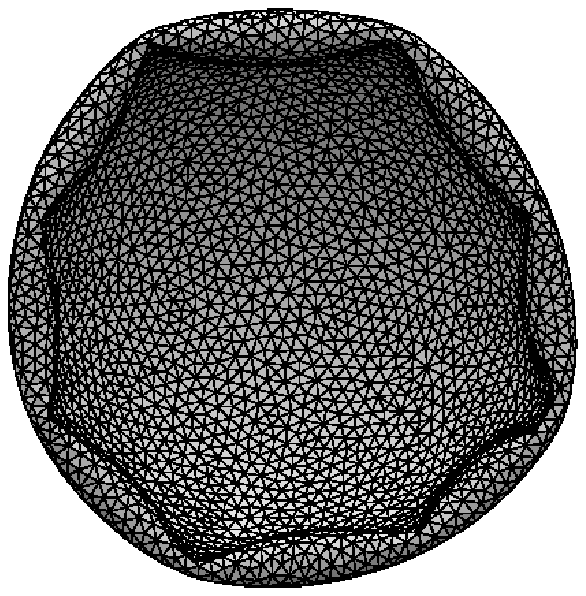}

\caption{\label{fig:ExpVsSimu}(a) Capsule obtained from evaporation in air
of a spherical shell of mean radius 870 nm, shell thickness 150 nm
(TEM image). (b) Simulation obtained for $\gamma$=271 and $\nu$=0.333
(hence equivalent to $\frac{d}{R}=0.172$), $\frac{\Delta V}{V}$=0.854,
3594 vertices. (c) Simulation: $\gamma$=2333 and $\nu$=0.333 (equivalent
to $\frac{d}{R}=0.0586$), $\frac{\Delta V}{V}$=0.854, 3594 vertices.
This conformation is to be compared with figure \ref{fig:Spherical-shells-buckled}c
(d) Simulation:$\gamma$=20995 and $\nu$=0.333(equivalent to $\frac{d}{R}=0.0195$),
$\frac{\Delta V}{V}$=0.854, 3594 vertices. This conformation is to
be compared with figure \ref{fig:Spherical-shells-buckled}e.}
\end{figure}

Figures \ref{fig:ExpVsSimu}-a and \ref{fig:ExpVsSimu}-b shows that
we could accurately reproduce the shape of axisymmetric capsules.
In the simulation displayed here, we took $\gamma$=271 and $\nu$=0.333,
which corresponds to $d/R$=0.172. 

For wrinkled bowls (examples displayed on figures \ref{fig:ExpVsSimu}-c
and \ref{fig:ExpVsSimu}-d), the conformations obtained just before
interpenetration are also very similar to shapes observed experimentally
(fig. \ref{fig:Spherical-shells-buckled}-c and \ref{fig:Spherical-shells-buckled}-e).
Like in the experiments, wrinkles do appear for thinner shells. Furthermore,
the parameters for which simulations provide wrinkles are consistent
with the shells' characteristics: $\sqrt{\frac{12}{\gamma}\left(1-\frac{2\nu^{2}}{1-\nu}\right)}$=0.02
to 0.08, to be compared to the experimental value $\frac{d}{R}$=0.003
to 0.04.

\begin{figure}
\includegraphics[width=33mm,keepaspectratio]{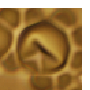}\includegraphics[width=22mm,keepaspectratio]{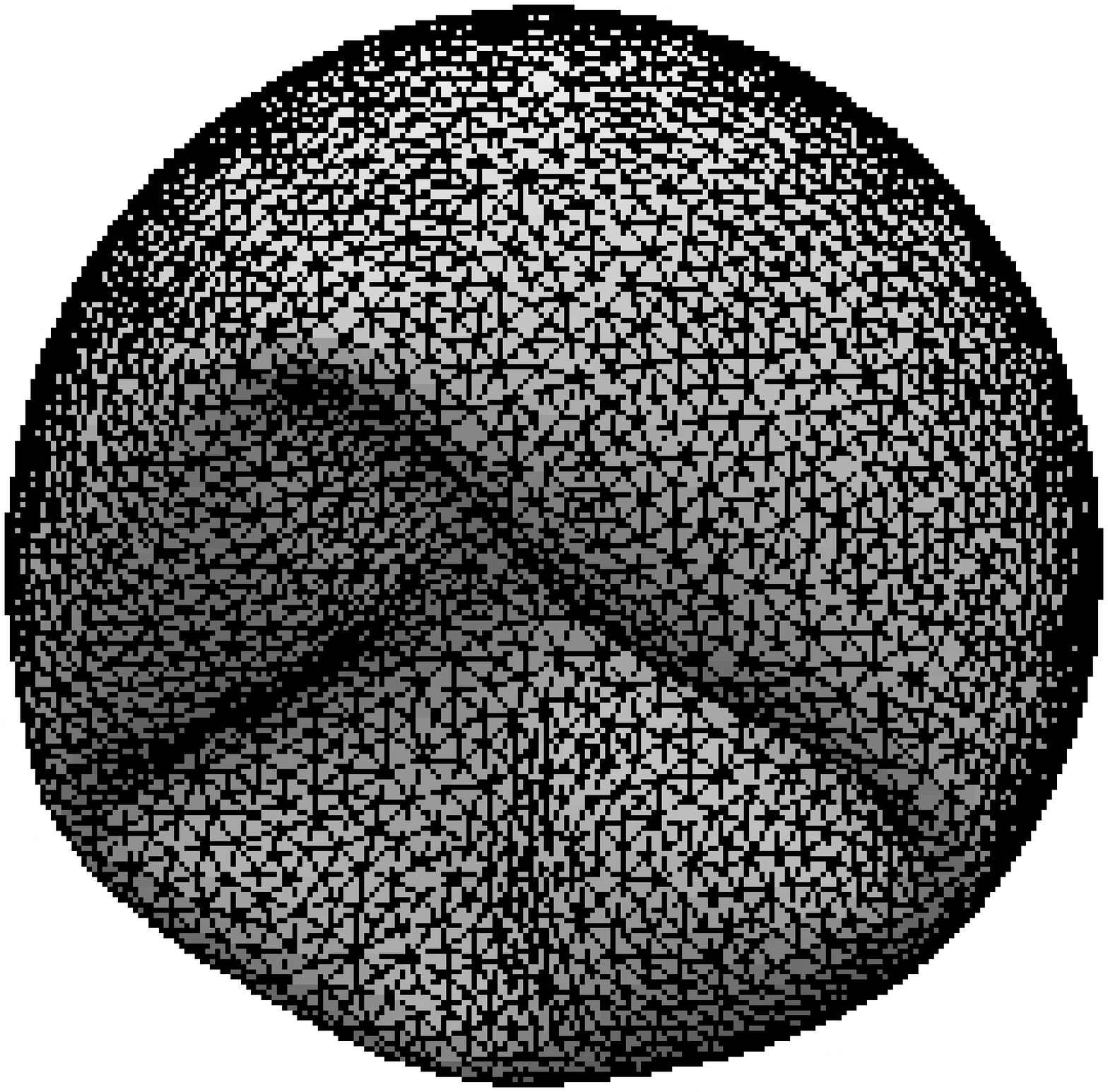}

\caption{\label{fig:Sym4}(a) Shell enclosing oil, evaporated in air ($d/R\approx0.012$).
Transmission optical microscopy, size 17 $\mu m\times\mu m$. (b)
Simulation obtained for $\gamma$=2916 and $\nu$=0.333 (equivalent
to $\frac{d}{R}$=0.064), by increasing the volume from $\frac{\Delta V}{V}$=0.474
to $\frac{\Delta V}{V}$=0.209, 3659 vertices.}
\end{figure}

Other experimental structures could also be reproduced by simulations,
such as the one displayed on figure \ref{fig:Sym4}a. This latter
was obtained through evaporation in air of a shell still containing
some of its inner oil (note: in this case the ethanol was added in
the aqueous solution later than usual, which possibly allowed polymerization
of longer chains \cite{Zoldesi2006} that cannot be dissolved by ethanol
\cite{NgLee2003}). Here the inner volume obviously does not correspond
to shell self -contact. In such a process, a shrinkage of the shell
itself when the water has fully evaporated can be invoked to explain
a behaviour comparable to a volume increase, such as in the simulation
of figure \ref{fig:Sym4}. The conformation, in this case, is stabilized
by oil-air interfaces.

All these results show that there is no need to invoke shell heterogeneity
to explain the shapes experimentally observed: bending and in-plane
stretching elasticity suffices. Next section provides more quantitative
insights on the simulation of elastic buckling.

\subsection{Quantitative comparison between simulations and elastic theoretical
calculations.\label{Simus_vs_theo}}

\begin{figure}
\includegraphics[scale=0.6]{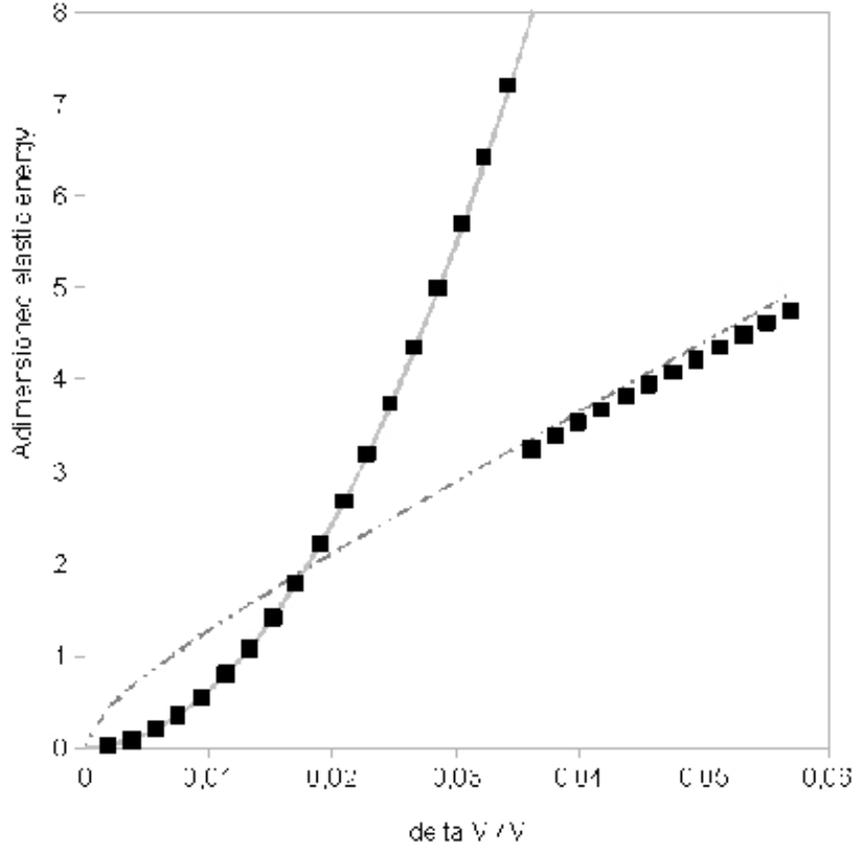}

\caption{\label{fig:Elastic-energy}Elastic energy, adimensioned by $\kappa$,
of a shrinking shell initially unstrained ($c_{0}=1/R$). Black squares:
result of Surface Evolver simulations, with $\gamma$=2916 and $\nu$=0.333.
Continuous line: normalized elastic energy $U_{sph}/\kappa=\gamma U_{sph}/(AR^{2})$
of a shrinking spherical shell, calculated with same $\gamma$ and
$\nu$ according to equation \ref{eq:Usphere}. Interrupted line:
adimensioned elastic energy $U_{axisym}/\kappa$ of a capsule (deformation
with a single axisymmetric depression), calculated with same$\gamma$
and $\nu$ according to equations \ref{eq:Ucapsule} and \ref{eq:deltaV_vs_alpha}.}
\end{figure}

The software Surface Evolver used to perform simulations provide the
elastic energy of each conformation. For the two conformations {}``sphere''
and {}``axisymmetric depression'', we compared this elastic energy
with the theoretical expressions $U_{sphere}$ (equation \ref{eq:Usphere})
and $U_{axisym}$ (equations \ref{eq:Ucapsule} and \ref{eq:deltaV_vs_alpha}).
The numerical data are very well fitted by the theory, as shown on
figure \ref{fig:Elastic-energy}. One sees that this first buckling
from a spherical shape to a conformation with a single axisymmetric
depression occurs with some hysteresis, \emph{i.e.} for volume variations
higher than the one corresponding to $U_{sphere}=U_{axisym}$.

\begin{figure}
\includegraphics[scale=0.7]{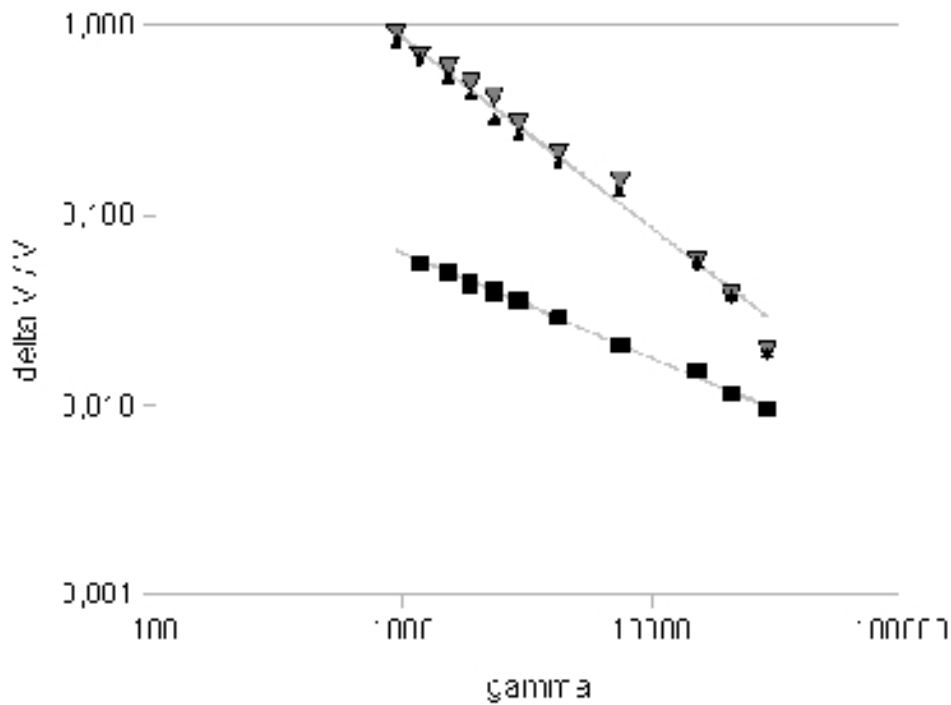}

\caption{\label{fig:BucklingOccurence}Primary buckling (black squares) from
sphere to capsule: relative volume at which the axisymmetric depression
appears in the simulations, for different $\gamma$. Interpolating
straight line : $\frac{\Delta V}{V}=2.8\,\gamma^{-0.55}$ . Secondary
buckling: polygonization of the circular rim (black upwards pointing
triangles) and apparition of wrinkles (gray downwards pointing triangles).
Interpolating straight line: $\frac{\Delta V}{V}=850\,\gamma^{-1}$.}
\end{figure}

Figure \ref{fig:BucklingOccurence} presents the buckling occurences
as a function of $\gamma$. The first buckling is determined without
ambiguity, as it is obvious from figure \ref{fig:Elastic-energy}.
The occurence of the second buckling, from an axisymmetric capsule
to a {}``wrinkled bowl'' conformation, is less easy to detect since
it corresponds neither to a discontinuity nor a singularity in energy.
We detected in fact two caracteristic values for $\frac{\Delta V}{V}$,
by visual observation of the conformations: the first one corresponds
to the loss of the axisymmetry, when the rim of the depression becomes
polygonal. Then the apices of the polygon becomes sharper (they tend
to form the extremity of a \emph{d}-cone \cite{Pauchard1998,Chaïeb1998}),
and the inner part of the rim becomes convex between two successive
apices: at this point we consider that the conformation holds wrinkles,
and this second {}``threshold'' is recorded. Figure \ref{fig:BucklingOccurence}
shows that both values are quite close and present the same power-law
in $\gamma^{-1}$. Extrapolation intercepts with $\frac{\Delta V}{V}$=1
at a value $\gamma_{c}$=850, which is consistent with our simulations
showing that secondary buckling appears for $\gamma$ between 583
and 933 ($\frac{d}{R}\approx0.1$). This can be compared with calculations
of ref \cite{Fitch68} that forecast a threshold $\gamma$ of 1345
for the apparition of wrinkles on a clamped cap submitted to concentrated
load.

The first buckling (sphere toward axisymmetric capsule) happens for
thresholds values of the relative volume variation that vary in a
power-law with the Föppl-von Karman parameter: $\frac{\Delta V}{V}\propto\gamma^{-0.55}$.
Despite the slight hysteresis in the primary buckling, this exponent
is very close to the $-0.6$ theoretically proposed in equation \ref{eq:PowerLawBuckling2}.

\section{Discussion.\label{sec:Discussion.}}

Experimental and numerical results showed that wrinkled bowls are
preferentially observed when a very thin spherical shells lowers its
volume. This conformation is quite different from the structures (typically
discocytes or stomatocytes) usually obtained for vesicles, where the
in-plane elasticity is liquid-like (related parameter: 2D compressibility)\cite{Seifert1997}.
Wrinkles evidences the 2D-solid nature of the shells, since it is
needed to accomodate the surface of one hemisphere within the other
hemisphere, without an excessive cost in stretch energy.

It is interesting to note that a structure presenting 3 wrinkles had
been obtained by Lim \emph{et al.}\cite{Lim2002} in simulations of
red blood cells where the elastic properties of respectively the cytoskeleton
and an homogeneous asymmetrical phospholipid bilayer were included
in a similar numerical model with elastic bending, spontaneous curvature
and stretching\cite{Mukhopad2002}. But the bending/stretching ratio
in these biological objects, where bending and stretching have different
origin, was higher than the range of similar values for a thin shell
of isotropic material. This probably prevented these authors from
obtaining shapes with more wrinkles.

\begin{figure}
\includegraphics[scale=0.15]{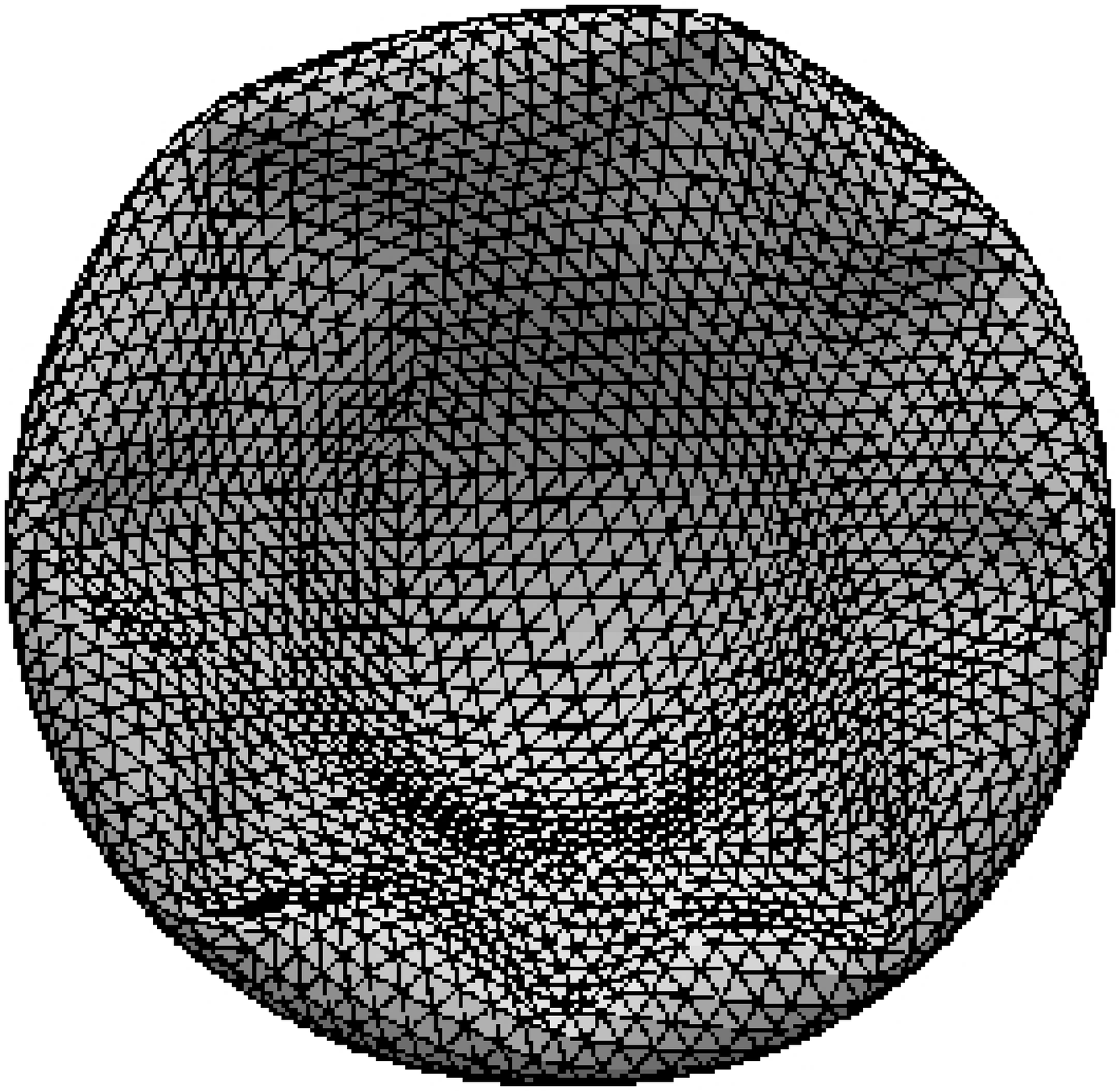}\includegraphics[scale=0.25]{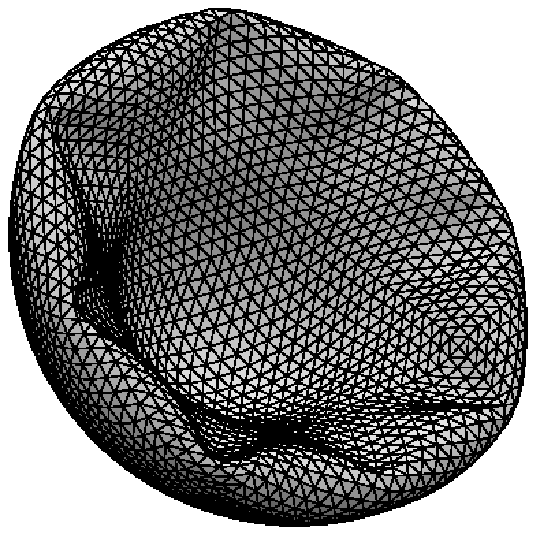}

\caption{\label{fig:Soleil}Simulation with $\gamma$=15163, $\nu$=0.333
and $\frac{\Delta V}{V}$=0.645. Such unusual conformation can be
obtained from a capsule at lower $\gamma$; its energy is 1.42 times
lower than the configuration with wrinkles ($W$=7) obtained through
progressive volume decrease for similar parameters.}
\end{figure}

Besides, the simulations presented here do not necessarily provide
the energies of lowest configuration. As an example, we could, by
following another path in the phase diagram of elastic and geometric
parameters, obtain a totally new conformation of much lower energy
than the wrinkled bowls (figure \ref{fig:Soleil}). But this conformation
very likely corresponds to an energy trough too narrow to have been
encountered in our experimental situation. Anyway, we are not looking
for equilibrium configurations: we are trying to understand what really
happens when a colloidal shell shrinks. It is well-known that many
buckling conformations can be quenched in non-absolute energy minima.
Our study, putting into evidence qualitative as well as quantitative
convergences between experiments, theory and simulations, strongly
suggests that our simulations with a progressive decrease of the inner
volume can reproduce the path followed by the buckling of real shells.
The shapes observed are compatible both with self-contact, which would
explain their stabilization, and with shell homogeneity. Besides,
the conformation (and furthermore the number of wrinkles) gives an
indication on the shell relative thickness range.

Structures comparable to wrinkled bowls have already been observed
experimentally on millimetric half-spheres submitted to a localized\cite{Pauchard1998,Kitching75}
or a planar\cite{Cui2004} load, theoretically forecasted\cite{Fitch68,Kitching75},
or numerically obtained by simulation of a sphere adhering on a flat
surface\cite{Komura2005}, but here we did put in evidence that such
structures can also be obtained with an isotropic force distribution.

\section{Conclusion}

Non-trivial buckled shapes were obtained by evaporating or dissolving
the solvent enclosed in porous colloidal shells.

We have shown that the deformations of such objects are consistent
with a model of homogeneous thin spherical shells with bending and
in-plane stretching elasticity submitted to an isotropic external
pressure. The numerical simulations showed that a primary buckling
leading to capsules (holding a single axisymmetric depression) can
be followed by a secondary buckling where the depression wrinkles.
This happens for decreasing volume variations when the relative thickness
of the shell is reduced, and the number of wrinkles concomitantly
increases. Simulations and experiments qualitatively and quantitatively
confirm each other.

These new results suggest that evaporation or dissolution of inner
solvent is a promising way to obtain, from a monodisperse enough population
of colloids, a monodisperse suspension of anisotropic objects with
geometric parameters tunable through the characteristics of the initial
spherical shell.

\subsection*{Acknowledgements :}

We thank L. Pauchard and N. Tsapis for their interest in this work,
P. Peyla for elasticity discussions, S. J. Cox for his patient help
in debugging SE subroutines, and K. Brakke for supplying and maintaining
the free software Surface Evolver with impressive swiftness. This
work was partially supported by the Deutsche Forschungsgemeinschaft
(DFG) within the SFB-TR6 program {}``Physics of colloidal dispersions
in external fields'', and by the D. G. A. (Direction Générale de
l'Armement).

\end{document}